%
%
%
%
%
%
%
%
%
%
%
%
%
%
%
%
\catcode`\@=11 
\let\rel@x=\relax
\let\n@expand=\relax
\def\pr@tect{\let\n@expand=\noexpand}
\let\protect=\pr@tect
\let\gl@bal=\global
%
%
%
\newfam\cpfam
\newdimen\b@gheight             \b@gheight=12pt
\newcount\f@ntkey               \f@ntkey=0
\def\f@m{\afterassignment\samef@nt\f@ntkey=}
\def\samef@nt{\fam=\f@ntkey \the\textfont\f@ntkey\rel@x}
\def\setstr@t{\setbox\strutbox=\hbox{\vrule height 0.85\b@gheight
                                depth 0.35\b@gheight width\z@ }}
%
%
%
%
%

\font\fourteenrm  =cmr12 scaled\magstep1
\font\twelverm    =cmr12
\font\ninerm      =cmr9
\font\sixrm       =cmr6

\font\fourteenbf  =cmbx12 scaled\magstep1
\font\twelvebf    =cmbx12
\font\ninebf      =cmbx9
\font\sixbf       =cmbx6
\font\seventeeni  =cmmi12 scaled\magstep2    \skewchar\seventeeni='177
\font\fourteeni   =cmmi12 scaled\magstep1     \skewchar\fourteeni='177
\font\twelvei     =cmmi12                       \skewchar\twelvei='177
\font\ninei       =cmmi9                          \skewchar\ninei='177
\font\sixi        =cmmi6                           \skewchar\sixi='177
\font\seventeensy =cmsy10 scaled\magstep3    \skewchar\seventeensy='60
\font\fourteensy  =cmsy10 scaled\magstep2     \skewchar\fourteensy='60
\font\twelvesy    =cmsy10 scaled\magstep1       \skewchar\twelvesy='60
\font\ninesy      =cmsy9                          \skewchar\ninesy='60
\font\sixsy       =cmsy6                           \skewchar\sixsy='60

\font\fourteenex  =cmex10 scaled\magstep2
\font\twelveex    =cmex10 scaled\magstep1

\font\fourteensl  =cmsl12 scaled\magstep1
\font\twelvesl    =cmsl12
\font\ninesl      =cmsl9

\font\fourteenit  =cmti12 scaled\magstep1
\font\twelveit    =cmti12
\font\nineit      =cmti9
\font\fourteentt  =cmtt12 scaled\magstep1
\font\twelvett    =cmtt12
\font\fourteencp  =cmcsc10 scaled\magstep2
\font\twelvecp    =cmcsc10 scaled\magstep1
\font\tencp       =cmcsc10
%
%
\def\fourteenf@nts{\relax
    \textfont0=\fourteenrm          \scriptfont0=\tenrm
      \scriptscriptfont0=\sevenrm
    \textfont1=\fourteeni           \scriptfont1=\teni
      \scriptscriptfont1=\seveni
    \textfont2=\fourteensy          \scriptfont2=\tensy
      \scriptscriptfont2=\sevensy
    \textfont3=\fourteenex          \scriptfont3=\twelveex
      \scriptscriptfont3=\tenex
    \textfont\itfam=\fourteenit     \scriptfont\itfam=\tenit
    \textfont\slfam=\fourteensl     \scriptfont\slfam=\tensl
    \textfont\bffam=\fourteenbf     \scriptfont\bffam=\tenbf
      \scriptscriptfont\bffam=\sevenbf
    \textfont\ttfam=\fourteentt
    \textfont\cpfam=\fourteencp }
\def\twelvef@nts{\relax
    \textfont0=\twelverm          \scriptfont0=\ninerm
      \scriptscriptfont0=\sixrm
    \textfont1=\twelvei           \scriptfont1=\ninei
      \scriptscriptfont1=\sixi
    \textfont2=\twelvesy           \scriptfont2=\ninesy
      \scriptscriptfont2=\sixsy
    \textfont3=\twelveex          \scriptfont3=\tenex
      \scriptscriptfont3=\tenex
    \textfont\itfam=\twelveit     \scriptfont\itfam=\nineit
    \textfont\slfam=\twelvesl     \scriptfont\slfam=\ninesl
    \textfont\bffam=\twelvebf     \scriptfont\bffam=\ninebf
      \scriptscriptfont\bffam=\sixbf
    \textfont\ttfam=\twelvett
    \textfont\cpfam=\twelvecp }
\def\tenf@nts{\relax
    \textfont0=\tenrm          \scriptfont0=\sevenrm
      \scriptscriptfont0=\fiverm
    \textfont1=\teni           \scriptfont1=\seveni
      \scriptscriptfont1=\fivei
    \textfont2=\tensy          \scriptfont2=\sevensy
      \scriptscriptfont2=\fivesy
    \textfont3=\tenex          \scriptfont3=\tenex
      \scriptscriptfont3=\tenex
    \textfont\itfam=\tenit     \scriptfont\itfam=\seveni  
    \textfont\slfam=\tensl     \scriptfont\slfam=\sevenrm 
    \textfont\bffam=\tenbf     \scriptfont\bffam=\sevenbf
      \scriptscriptfont\bffam=\fivebf
    \textfont\ttfam=\tentt
    \textfont\cpfam=\tencp }
\def\rm{\n@expand\f@m0 }
\def\mit{\n@expand\f@m1 }         
\def\cal{\n@expand\f@m2 }
\def\it{\n@expand\f@m\itfam}
\def\sl{\n@expand\f@m\slfam}
\def\bf{\n@expand\f@m\bffam}
\def\tt{\n@expand\f@m\ttfam}
\def\caps{\n@expand\f@m\cpfam}    
\def\em@{\rel@x\ifnum\f@ntkey=0 \it \else
        \ifnum\f@ntkey=\bffam \it \else \rm \fi \fi }
\def\em{\n@expand\em@}
\def\fourteenpoint{\fourteenf@nts \samef@nt \b@gheight=14pt \setstr@t }
\def\twelvepoint{\twelvef@nts \samef@nt \b@gheight=12pt \setstr@t }
\def\tenpoint{\tenf@nts \samef@nt \b@gheight=10pt \setstr@t }
\normalbaselineskip = 20pt plus 0.2pt minus 0.1pt
\normallineskip = 1.5pt plus 0.1pt minus 0.1pt
\normallineskiplimit = 1.5pt
\newskip\normaldisplayskip
\normaldisplayskip = 20pt plus 5pt minus 10pt
\newskip\normaldispshortskip
\normaldispshortskip = 6pt plus 5pt
\newskip\normalparskip
\normalparskip = 6pt plus 2pt minus 1pt
\newskip\skipregister
\skipregister = 5pt plus 2pt minus 1.5pt
\newif\ifsingl@
\newif\ifdoubl@
\newif\iftwelv@  \twelv@true
\def\singlespace{\singl@true\doubl@false\spaces@t}
\def\doublespace{\singl@false\doubl@true\spaces@t}
\def\normalspace{\singl@false\doubl@false\spaces@t}
\def\Tenpoint{\tenpoint\twelv@false\spaces@t}
\def\Twelvepoint{\twelvepoint\twelv@true\spaces@t}
\def\spaces@t{\rel@x
      \iftwelv@ \ifsingl@\subspaces@t3:4;\else\subspaces@t1:1;\fi
       \else \ifsingl@\subspaces@t3:5;\else\subspaces@t4:5;\fi \fi
      \ifdoubl@ \multiply\baselineskip by 5
         \divide\baselineskip by 4 \fi }
\def\subspaces@t#1:#2;{
      \baselineskip = \normalbaselineskip
      \multiply\baselineskip by #1 \divide\baselineskip by #2
      \lineskip = \normallineskip
      \multiply\lineskip by #1 \divide\lineskip by #2
      \lineskiplimit = \normallineskiplimit
      \multiply\lineskiplimit by #1 \divide\lineskiplimit by #2
      \parskip = \normalparskip
      \multiply\parskip by #1 \divide\parskip by #2
      \abovedisplayskip = \normaldisplayskip
      \multiply\abovedisplayskip by #1 \divide\abovedisplayskip by #2
      \belowdisplayskip = \abovedisplayskip
      \abovedisplayshortskip = \normaldispshortskip
      \multiply\abovedisplayshortskip by #1
        \divide\abovedisplayshortskip by #2
      \belowdisplayshortskip = \abovedisplayshortskip
      \advance\belowdisplayshortskip by \belowdisplayskip
      \divide\belowdisplayshortskip by 2
      \smallskipamount = \skipregister
      \multiply\smallskipamount by #1 \divide\smallskipamount by #2
      \medskipamount = \smallskipamount \multiply\medskipamount by 2
      \bigskipamount = \smallskipamount \multiply\bigskipamount by 4 }
\def\normalbaselines{ \baselineskip=\normalbaselineskip
   \lineskip=\normallineskip \lineskiplimit=\normallineskip
   \iftwelv@\else \multiply\baselineskip by 4 \divide\baselineskip by 5
     \multiply\lineskiplimit by 4 \divide\lineskiplimit by 5
     \multiply\lineskip by 4 \divide\lineskip by 5 \fi }
\Twelvepoint  
\interlinepenalty=50
\interfootnotelinepenalty=5000
\predisplaypenalty=9000
\postdisplaypenalty=500
\hfuzz=1pt
\vfuzz=0.2pt
\newdimen\HOFFSET  \HOFFSET=0pt
\newdimen\VOFFSET  \VOFFSET=0pt
\newdimen\HSWING   \HSWING=0pt
\dimen\footins=8in
%
%
%
\newskip\pagebottomfiller
\pagebottomfiller=\z@ plus \z@ minus \z@
\def\pagecontents{
   \ifvoid\topins\else\unvbox\topins\vskip\skip\topins\fi
   \dimen@ = \dp255 \unvbox255
   \vskip\pagebottomfiller
   \ifvoid\footins\else\vskip\skip\footins\footrule\unvbox\footins\fi
   \ifr@ggedbottom \kern-\dimen@ \vfil \fi }
\def\makeheadline{\vbox to 0pt{ \skip@=\topskip
      \advance\skip@ by -12pt \advance\skip@ by -2\normalbaselineskip
      \vskip\skip@ \line{\vbox to 12pt{}\the\headline} \vss
      }\nointerlineskip}
\def\makefootline{\baselineskip = 1.5\normalbaselineskip
                 \line{\the\footline}}
\newif\iffrontpage
\newif\ifp@genum
\def\nopagenumbers{\p@genumfalse}
\def\pagenumbers{\p@genumtrue}
\pagenumbers
\newtoks\paperheadline
\newtoks\paperfootline
\newtoks\letterheadline
\newtoks\letterfootline
\newtoks\letterinfo
\newtoks\date
\paperheadline={\hfil}
\paperfootline={\hss\iffrontpage\else\ifp@genum\tenrm\folio\hss\fi\fi}
\letterheadline{\iffrontpage \hfil \else
    \rm \ifp@genum page~~\folio\fi \hfil\the\date \fi}
\letterfootline={\iffrontpage\the\letterinfo\else\hfil\fi}
\letterinfo={\hfil}
\def\monthname{\rel@x\ifcase\month 0/\or January\or February\or
   March\or April\or May\or June\or July\or August\or September\or
   October\or November\or December\else\number\month/\fi}
\def\today{\monthname~\number\day, \number\year}
\date={\today}
\headline=\paperheadline 
\footline=\paperfootline 
\countdef\pageno=1      \countdef\pagen@=0
\countdef\pagenumber=1  \pagenumber=1
\def\advancepageno{\gl@bal\advance\pagen@ by 1
   \ifnum\pagenumber<0 \gl@bal\advance\pagenumber by -1
    \else\gl@bal\advance\pagenumber by 1 \fi
    \gl@bal\frontpagefalse  \swing@ }
\def\folio{\ifnum\pagenumber<0 \romannumeral-\pagenumber
           \else \number\pagenumber \fi }
\def\swing@{\ifodd\pagenumber \gl@bal\advance\hoffset by -\HSWING
             \else \gl@bal\advance\hoffset by \HSWING \fi }
\def\footrule{\dimen@=\prevdepth\nointerlineskip
   \vbox to 0pt{\vskip -0.25\baselineskip \hrule width 0.35\hsize \vss}
   \prevdepth=\dimen@ }
\let\footnotespecial=\rel@x
\newdimen\footindent
\footindent=24pt
\def\Textindent#1{\noindent\llap{#1\enspace}\ignorespaces}
\def\Vfootnote#1{\insert\footins\bgroup
   \interlinepenalty=\interfootnotelinepenalty \floatingpenalty=20000
   \singl@true\doubl@false\Tenpoint
   \splittopskip=\ht\strutbox \boxmaxdepth=\dp\strutbox
   \leftskip=\footindent \rightskip=\z@skip
   \parindent=0.5\footindent \parfillskip=0pt plus 1fil
   \spaceskip=\z@skip \xspaceskip=\z@skip \footnotespecial
   \Textindent{#1}\footstrut\futurelet\next\fo@t}

\def\vfootnote#1{\Vfootnote{${#1}$}}
\def\footnote#1{\attach{#1}\vfootnote{#1}}

\def\foot{\attach\footsymbolgen\vfootnote{\footsymbol}}
\let\footsymbol=\star
\newcount\lastf@@t           \lastf@@t=-1
\newcount\footsymbolcount    \footsymbolcount=0
\newif\ifPhysRev
\def\footsymbolgen{\bumpfootsymbolcount \generatefootsymbol \footsymbol }
\def\bumpfootsymbolcount{\rel@x
   \iffrontpage \bumpfootsymbolpos \else \advance\lastf@@t by 1
     \ifPhysRev \bumpfootsymbolneg \else \bumpfootsymbolpos \fi \fi
   \gl@bal\lastf@@t=\pagen@ }
\def\bumpfootsymbolpos{\ifnum\footsymbolcount <0
                            \gl@bal\footsymbolcount =0 \fi
    \ifnum\lastf@@t<\pagen@ \gl@bal\footsymbolcount=0
     \else \gl@bal\advance\footsymbolcount by 1 \fi }
\def\bumpfootsymbolneg{\ifnum\footsymbolcount >0
             \gl@bal\footsymbolcount =0 \fi
         \gl@bal\advance\footsymbolcount by -1 }
\def\fd@f#1 {\xdef\footsymbol{\mathchar"#1 }}
\def\generatefootsymbol{\ifcase\footsymbolcount \fd@f 13F \or \fd@f 279
        \or \fd@f 27A \or \fd@f 278 \or \fd@f 27B \else
        \ifnum\footsymbolcount <0 \fd@f{023 \number-\footsymbolcount }
         \else \fd@f 203 {\loop \ifnum\footsymbolcount >5
                \fd@f{203 \footsymbol } \advance\footsymbolcount by -1
                \repeat }\fi \fi }

\def\nonfrenchspacing{\sfcode`\.=3001 \sfcode`\!=3000 \sfcode`\?=3000
        \sfcode`\:=2000 \sfcode`\;=1500 \sfcode`\,=1251 }
\nonfrenchspacing
\newdimen\d@twidth
{\setbox0=\hbox{s.} \gl@bal\d@twidth=\wd0 \setbox0=\hbox{s}
        \gl@bal\advance\d@twidth by -\wd0 }
\def\removehglue{\loop \unskip \ifdim\lastskip >\z@ \repeat }
\def\roll@ver#1{\removehglue \nobreak \count255 =\spacefactor \dimen@=\z@
        \ifnum\count255 =3001 \dimen@=\d@twidth \fi
        \ifnum\count255 =1251 \dimen@=\d@twidth \fi
    \iftwelv@ \kern-\dimen@ \else \kern-0.83\dimen@ \fi
   #1\spacefactor=\count255 }
\def\step@ver#1{\rel@x \ifmmode #1\else \ifhmode
        \roll@ver{${}#1$}\else {\setbox0=\hbox{${}#1$}}\fi\fi }
\def\attach#1{\step@ver{\strut^{\mkern 2mu #1} }}
%
%
%
\newcount\chapternumber      \chapternumber=0
\newcount\sectionnumber      \sectionnumber=0
\newcount\equanumber         \equanumber=0
\let\chapterlabel=\rel@x
\let\sectionlabel=\rel@x
\newtoks\chapterstyle        \chapterstyle={\Number}
\newtoks\sectionstyle        \sectionstyle={\Number}
\newskip\chapterskip         \chapterskip=\bigskipamount
\newskip\sectionskip         \sectionskip=\medskipamount
\newskip\headskip            \headskip=8pt plus 3pt minus 3pt
\newdimen\chapterminspace    \chapterminspace=15pc
\newdimen\sectionminspace    \sectionminspace=10pc
\newdimen\referenceminspace  \referenceminspace=20pc
\newif\ifcn@                 \cn@true
\newif\ifcn@@                \cn@@false
\def\numberedchapters{\cn@true}
\def\unnumberedchapters{\cn@false\sequentialequations}
\def\chapterreset{\gl@bal\advance\chapternumber by 1
   \ifnum\equanumber<0 \else\gl@bal\equanumber=0\fi
   \sectionnumber=0 \let\sectionlabel=\rel@x
   \ifcn@ \gl@bal\cn@@true {\pr@tect
       \xdef\chapterlabel{\the\chapterstyle{\the\chapternumber}}}%
    \else \gl@bal\cn@@false \gdef\chapterlabel{\rel@x}\fi }
\def\@alpha#1{\count255='140 \advance\count255 by #1\char\count255}
 \def\alphabetic{\n@expand\@alpha}
\def\@Alpha#1{\count255='100 \advance\count255 by #1\char\count255}
 \def\Alphabetic{\n@expand\@Alpha}
\def\@Roman#1{\uppercase\expandafter{\romannumeral #1}}
 \def\Roman{\n@expand\@Roman}
\def\@roman#1{\romannumeral #1}    \def\roman{\n@expand\@roman}
\def\@number#1{\number #1}         \def\Number{\n@expand\@number}
\def\BLANK#1{\rel@x}               
\def\titleparagraphs{\interlinepenalty=9999
     \leftskip=0.03\hsize plus 0.22\hsize minus 0.03\hsize
     \rightskip=\leftskip \parfillskip=0pt
     \hyphenpenalty=9000 \exhyphenpenalty=9000
     \tolerance=9999 \pretolerance=9000
     \spaceskip=0.333em \xspaceskip=0.5em }
\def\titlestyle#1{\par\begingroup \titleparagraphs
     \iftwelv@\fourteenpoint\else\twelvepoint\fi
   \noindent #1\par\endgroup }
\def\spacecheck#1{\dimen@=\pagegoal\advance\dimen@ by -\pagetotal
   \ifdim\dimen@<#1 \ifdim\dimen@>0pt \vfil\break \fi\fi}
\def\chapter#1{\par \penalty-300 \vskip\chapterskip
   \spacecheck\chapterminspace
   \chapterreset \titlestyle{\ifcn@@\chapterlabel.~\fi #1}
   \nobreak\vskip\headskip \penalty 30000
   {\pr@tect\wlog{\string\chapter\space \chapterlabel}} }

\def\section#1{\par \ifnum\lastpenalty=30000\else
   \penalty-200\vskip\sectionskip \spacecheck\sectionminspace\fi
   \gl@bal\advance\sectionnumber by 1
   {\pr@tect
   \xdef\sectionlabel{\ifcn@@ \chapterlabel.\fi
       \the\sectionstyle{\the\sectionnumber}}%
   \wlog{\string\section\space \sectionlabel}}%
   \noindent {\caps\enspace\sectionlabel.~~#1}\par
   \nobreak\vskip\headskip \penalty 30000 }
\def\subsection#1{\par
   \ifnum\the\lastpenalty=30000\else \penalty-100\smallskip \fi
   \noindent\undertext{#1}\enspace \vadjust{\penalty5000}}

\def\undertext#1{\vtop{\hbox{#1}\kern 1pt \hrule}}

\def\ack{\subsection{Acknowledgements:}}
\def\APPENDIX#1#2{\par\penalty-300\vskip\chapterskip
   \spacecheck\chapterminspace \chapterreset \xdef\chapterlabel{#1}
   \titlestyle{APPENDIX #2} \nobreak\vskip\headskip \penalty 30000
   \wlog{\string\Appendix~\chapterlabel} }
\def\Appendix#1{\APPENDIX{#1}{#1}}
\def\appendix{\APPENDIX{A}{}}
%
%
%
\def\eqname#1{\rel@x {\pr@tect
  \ifnum\equanumber<0 \xdef#1{{\rm(\number-\equanumber)}}%
     \gl@bal\advance\equanumber by -1
  \else \gl@bal\advance\equanumber by 1
   \xdef#1{{\rm(\ifcn@@ \chapterlabel.\fi \number\equanumber)}}\fi
  }#1}

\def\eqn{\eqno\eqname}

\def\eqinsert#1{\noalign{\dimen@=\prevdepth \nointerlineskip
   \setbox0=\hbox to\displaywidth{\hfil #1}
   \vbox to 0pt{\kern 0.5\baselineskip\hbox{$\!\box0\!$}\vss}
   \prevdepth=\dimen@}}
%

%
%
\def\GENITEM#1;#2{\par \hangafter=0 \hangindent=#1
    \Textindent{$ #2 $}\ignorespaces}
\outer\def\newitem#1=#2;{\gdef#1{\GENITEM #2;}}

\newdimen\itemsize                \itemsize=30pt
\newitem\item=1\itemsize;
\newitem\sitem=1.75\itemsize;     
\newitem\ssitem=2.5\itemsize;     
\outer\def\newlist#1=#2&#3&#4;{\toks0={#2}\toks1={#3}%
   \count255=\escapechar \escapechar=-1
   \alloc@0\list\countdef\insc@unt\listcount     \listcount=0
   \edef#1{\par
      \countdef\listcount=\the\allocationnumber
      \advance\listcount by 1
      \hangafter=0 \hangindent=#4
      \Textindent{\the\toks0{\listcount}\the\toks1}}
   \expandafter\expandafter\expandafter
    \edef\c@t#1{begin}{\par
      \countdef\listcount=\the\allocationnumber \listcount=1
      \hangafter=0 \hangindent=#4
      \Textindent{\the\toks0{\listcount}\the\toks1}}
   \expandafter\expandafter\expandafter
    \edef\c@t#1{con}{\par \hangafter=0 \hangindent=#4 \noindent}
   \escapechar=\count255}
\def\c@t#1#2{\csname\string#1#2\endcsname}
\newlist\point=\Number&.&1.0\itemsize;
\newlist\subpoint=(\alphabetic&)&1.75\itemsize;
\newlist\subsubpoint=(\roman&)&2.5\itemsize;
%

%
%
%
%
\newcount\referencecount     \referencecount=0
\newcount\lastrefsbegincount \lastrefsbegincount=0
\newif\ifreferenceopen       \newwrite\referencewrite
\newdimen\refindent          \refindent=30pt
\def\normalrefmark#1{\attach{\scriptscriptstyle [ #1 ] }}
\let\PRrefmark=\attach
\def\NPrefmark#1{\step@ver{{\;[#1]}}}
\def\refmark#1{\rel@x\ifPhysRev\PRrefmark{#1}\else\normalrefmark{#1}\fi}
\def\refend@{\refmark{\number\referencecount}}
\def\refend{\refend@{}\space }
\def\refsend{\refmark{\count255=\referencecount
   \advance\count255 by-\lastrefsbegincount
   \ifcase\count255 \number\referencecount
   \or \number\lastrefsbegincount,\number\referencecount
   \else \number\lastrefsbegincount-\number\referencecount \fi}\space }
\def\REFNUM#1{\rel@x \gl@bal\advance\referencecount by 1
    \xdef#1{\the\referencecount }}
\def\Refnum#1{\REFNUM #1\refend@ } 
\def\REF#1{\REFNUM #1\R@FWRITE\ignorespaces}
\def\Ref#1{\Refnum #1\REFWRITE }
\def\ref{\Ref\?}
\def\REFS#1{\REFNUM #1\gl@bal\lastrefsbegincount=\referencecount
    \REFWRITE }

\def\r@fitem#1{\par \hangafter=0 \hangindent=\refindent \Textindent{#1}}
\def\refitem#1{\r@fitem{#1.}}
\def\NPrefitem#1{\r@fitem{[#1]}}
\def\NPrefs{\let\refmark=\NPrefmark \let\refitem=NPrefitem}
\def\REFWRITE{\R@FWRITE\rel@x }
\def\R@FWRITE#1{\ifreferenceopen \else \gl@bal\referenceopentrue
     \immediate\openout\referencewrite=\jobname.refs
     \toks@={\begingroup \refoutspecials \catcode`\^^M=10 }%
     \immediate\write\referencewrite{\the\toks@}\fi
    \immediate\write\referencewrite{\noexpand\refitem %
                                    {\the\referencecount}}%
    \p@rse@ndwrite \referencewrite #1}
\begingroup
 \catcode`\^^M=\active \let^^M=\relax %
 \gdef\p@rse@ndwrite#1#2{\begingroup \catcode`\^^M=12 \newlinechar=`\^^M%
         \chardef\rw@write=#1\sc@nlines#2}%
 \gdef\sc@nlines#1#2{\sc@n@line \g@rbage #2^^M\endsc@n \endgroup #1}%
 \gdef\sc@n@line#1^^M{\expandafter\toks@\expandafter{\deg@rbage #1}%
         \immediate\write\rw@write{\the\toks@}%
         \futurelet\n@xt \sc@ntest }%
\endgroup
\def\sc@ntest{\ifx\n@xt\endsc@n \let\n@xt=\rel@x
       \else \let\n@xt=\sc@n@notherline \fi \n@xt }
\def\sc@n@notherline{\sc@n@line \g@rbage }
\def\deg@rbage#1{}
\let\g@rbage=\relax    \let\endsc@n=\relax
\def\refout{\par\penalty-400\vskip\chapterskip
   \spacecheck\referenceminspace
   \ifreferenceopen \Closeout\referencewrite \referenceopenfalse \fi
   \line{\fourteenrm\hfil REFERENCES\hfil}\vskip\headskip
   \input \jobname.refs
   }
\def\refoutspecials{\sfcode`\.=1000 \interlinepenalty=1000
         \rightskip=\z@ plus 1em minus \z@ }
\def\Closeout#1{\toks0={\par\endgroup}\immediate\write#1{\the\toks0}%
   \immediate\closeout#1}
%
%
\newcount\figurecount     \figurecount=0
\newcount\tablecount      \tablecount=0
\newif\iffigureopen       \newwrite\figurewrite
\newif\iftableopen        \newwrite\tablewrite
\def\FIGNUM#1{\rel@x \gl@bal\advance\figurecount by 1
    \xdef#1{\the\figurecount}}
\def\FIGURE#1{\FIGNUM #1\F@GWRITE\ignorespaces }

\def\figitem#1{\r@fitem{#1)}}
\def\FIGWRITE{\F@GWRITE\rel@x }
\def\TABNUM#1{\rel@x \gl@bal\advance\tablecount by 1
    \xdef#1{\the\tablecount}}
\def\TABLE#1{\TABNUM #1\T@BWRITE\ignorespaces }

\def\tabitem#1{\r@fitem{#1:}}
\def\TABWRITE{\T@BWRITE\rel@x }
\def\F@GWRITE#1{\iffigureopen \else \gl@bal\figureopentrue
     \immediate\openout\figurewrite=\jobname.figs
     \toks@={\begingroup \catcode`\^^M=10 }%
     \immediate\write\figurewrite{\the\toks@}\fi
    \immediate\write\figurewrite{\noexpand\figitem %
                                 {\the\figurecount}}%
    \p@rse@ndwrite \figurewrite #1}
\def\T@BWRITE#1{\iftableopen \else \gl@bal\tableopentrue
     \immediate\openout\tablewrite=\jobname.tabs
     \toks@={\begingroup \catcode`\^^M=10 }%
     \immediate\write\tablewrite{\the\toks@}\fi
    \immediate\write\tablewrite{\noexpand\tabitem %
                                 {\the\tablecount}}%
    \p@rse@ndwrite \tablewrite #1}
\def\figout{\par\penalty-400
   \vskip\chapterskip\spacecheck\referenceminspace
   \iffigureopen \Closeout\figurewrite \figureopenfalse \fi
   \line{\fourteenrm\hfil FIGURE CAPTIONS\hfil}\vskip\headskip
   \input \jobname.figs
   }
\def\tabout{\par\penalty-400
   \vskip\chapterskip\spacecheck\referenceminspace
   \iftableopen \Closeout\tablewrite \tableopenfalse \fi
   \line{\fourteenrm\hfil TABLE CAPTIONS\hfil}\vskip\headskip
   \input \jobname.tabs
   }
%
%
%
\newbox\picturebox
\def\p@cht{\ht\picturebox }
\def\p@cwd{\wd\picturebox }
\def\p@cdp{\dp\picturebox }
\newdimen\xshift
\newdimen\yshift
\newdimen\captionwidth
\newskip\captionskip
\captionskip=15pt plus 5pt minus 3pt
\def\fullwidth{\captionwidth=\hsize }
\newtoks\Caption
\newif\ifcaptioned
\newif\ifselfcaptioned
\def\caption{\captionedtrue \Caption }
\newcount\linesabove
\newif\iffileexists
\newtoks\picfilename
\def\fil@#1 {\fileexiststrue \picfilename={#1}}
\def\file#1{\if=#1\let\n@xt=\fil@ \else \def\n@xt{\fil@ #1}\fi \n@xt }
\def\pl@t{\begingroup \pr@tect
    \setbox\picturebox=\hbox{}\fileexistsfalse
    \let\height=\p@cht \let\width=\p@cwd \let\depth=\p@cdp
    \xshift=\z@ \yshift=\z@ \captionwidth=\z@
    \Caption={}\captionedfalse
    \linesabove =0 \picturedefault }
\def\plot{\pl@t \selfcaptionedfalse }
\def\Picture#1{\gl@bal\advance\figurecount by 1
    \xdef#1{\the\figurecount}\pl@t \selfcaptionedtrue }

\def\s@vepicture{\iffileexists \parsefilename \redopicturebox \fi
   \ifdim\captionwidth>\z@ \else \captionwidth=\p@cwd \fi
   \xdef\lastpicture{\iffileexists
        \setbox0=\hbox{\raise\the\yshift \vbox{%
              \moveright\the\xshift\hbox{\picturedefinition}}}%
        \else \setbox0=\hbox{}\fi
         \ht0=\the\p@cht \wd0=\the\p@cwd \dp0=\the\p@cdp
         \vbox{\hsize=\the\captionwidth \line{\hss\box0 \hss }%
              \ifcaptioned \vskip\the\captionskip \noexpand\Tenpoint
                \ifselfcaptioned Figure~\the\figurecount.\enspace \fi
                \the\Caption \fi }}%
    \endgroup }
\let\endpicture=\s@vepicture
\def\savepicture#1{\s@vepicture \global\let#1=\lastpicture }
\def\displaypicture{\fullwidth \s@vepicture $$\lastpicture $${}}
\def\toppicture{\fullwidth \s@vepicture \topinsert
    \lastpicture \medskip \endinsert }
\def\midpicture{\fullwidth \s@vepicture \midinsert
    \lastpicture \endinsert }
%
%
\def\leftpicture{\pres@tpicture
    \dimen@i=\hsize \advance\dimen@i by -\dimen@ii
    \setbox\picturebox=\hbox to \hsize {\box0 \hss }%
    \wr@paround }
\def\rightpicture{\pres@tpicture
    \dimen@i=\z@
    \setbox\picturebox=\hbox to \hsize {\hss \box0 }%
    \wr@paround }
\def\pres@tpicture{\gl@bal\linesabove=\linesabove
    \s@vepicture \setbox\picturebox=\vbox{
         \kern \linesabove\baselineskip \kern 0.3\baselineskip
         \lastpicture \kern 0.3\baselineskip }%
    \dimen@=\p@cht \dimen@i=\dimen@
    \advance\dimen@i by \pagetotal
    \par \ifdim\dimen@i>\pagegoal \vfil\break \fi
    \dimen@ii=\hsize
    \advance\dimen@ii by -\parindent \advance\dimen@ii by -\p@cwd
    \setbox0=\vbox to\z@{\kern-\baselineskip \unvbox\picturebox \vss }}
\def\wr@paround{\Caption={}\count255=1
    \loop \ifnum \linesabove >0
         \advance\linesabove by -1 \advance\count255 by 1
         \advance\dimen@ by -\baselineskip
         \expandafter\Caption \expandafter{\the\Caption \z@ \hsize }%
      \repeat
    \loop \ifdim \dimen@ >\z@
         \advance\count255 by 1 \advance\dimen@ by -\baselineskip
         \expandafter\Caption \expandafter{%
             \the\Caption \dimen@i \dimen@ii }%
      \repeat
    \edef\n@xt{\parshape=\the\count255 \the\Caption \z@ \hsize }%
    \par\noindent \n@xt \strut \vadjust{\box\picturebox }}
\let\picturedefault=\relax
\let\parsefilename=\relax
\def\redopicturebox{\let\picturedefinition=\rel@x
   \errhelp=\disabledpictures
   \errmessage{This version of TeX cannot handle pictures.  Sorry.}}
\newhelp\disabledpictures
     {You will get a blank box in place of your picture.}
%
%
%
%
%
%
%
%
%
%
\def\FRONTPAGE{\ifvoid255\else\vfill\penalty-20000\fi
   \gl@bal\pagenumber=1     \gl@bal\chapternumber=0
   \gl@bal\equanumber=0     \gl@bal\sectionnumber=0
   \gl@bal\referencecount=0 \gl@bal\figurecount=0
   \gl@bal\tablecount=0     \gl@bal\frontpagetrue
   \gl@bal\lastf@@t=0       \gl@bal\footsymbolcount=0
   \gl@bal\cn@@false }

\def\papers{\papersize\headline=\paperheadline\footline=\paperfootline}
\def\papersize{\hsize=35pc \vsize=50pc \hoffset=0pc \voffset=1pc
   \advance\hoffset by\HOFFSET \advance\voffset by\VOFFSET
   \pagebottomfiller=0pc
   \skip\footins=\bigskipamount \normalspace }
\papers  
%
%
\newskip\lettertopskip       \lettertopskip=20pt plus 50pt
\newskip\letterbottomskip    \letterbottomskip=\z@ plus 100pt
\newskip\signatureskip       \signatureskip=40pt plus 3pt
\def\lettersize{\hsize=6.5in \vsize=8.5in \hoffset=0in \voffset=0.5in
   \advance\hoffset by\HOFFSET \advance\voffset by\VOFFSET
   \pagebottomfiller=\letterbottomskip
   \skip\footins=\smallskipamount \multiply\skip\footins by 3
   \singlespace }
\def\MEMO{\lettersize \headline=\letterheadline \footline={\hfil }%
   \let\rule=\memorule \FRONTPAGE \memohead }

\def\memodate{\afterassignment\MEMO \date }
\def\memit@m#1{\smallskip \hangafter=0 \hangindent=1in
    \Textindent{\caps #1}}
\def\subject{\memit@m{Subject:}}
\def\topic{\memit@m{Topic:}}
\def\from{\memit@m{From:}}
\def\to{\rel@x \ifmmode \rightarrow \else \memit@m{To:}\fi }
\def\memorule{\medskip\hrule height 1pt\bigskip}  
\def\memohead{\centerline{\fourteenrm MEMORANDUM}}
\newwrite\labelswrite
\newtoks\rw@toks
\def\letters{\lettersize
   \headline=\letterheadline \footline=\letterfootline
   \immediate\openout\labelswrite=\jobname.lab}

\let\letterhead=\rel@x
\def\addressee#1{\medskip\line{\hskip 0.75\hsize plus\z@ minus 0.25\hsize
                               \the\date \hfil }%
   \vskip \lettertopskip
   \ialign to\hsize{\strut ##\hfil\tabskip 0pt plus \hsize \crcr #1\crcr}
   \writelabel{#1}\medskip \noindent\hskip -\spaceskip \ignorespaces }
\def\rwl@begin#1\cr{\rw@toks={#1\crcr}\rel@x
   \immediate\write\labelswrite{\the\rw@toks}\futurelet\n@xt\rwl@next}
\def\rwl@next{\ifx\n@xt\rwl@end \let\n@xt=\rel@x
      \else \let\n@xt=\rwl@begin \fi \n@xt}
\let\rwl@end=\rel@x
\def\writelabel#1{\immediate\write\labelswrite{\noexpand\labelbegin}
     \rwl@begin #1\cr\rwl@end
     \immediate\write\labelswrite{\noexpand\labelend}}
\newtoks\FromAddress         \FromAddress={}
\newtoks\sendername          \sendername={}
\newbox\FromLabelBox
\newdimen\labelwidth          \labelwidth=6in
\def\makelabels{\afterassignment\Makelabels \sendername=}
\def\Makelabels{\FRONTPAGE \letterinfo={\hfil } \MakeFromBox
     \immediate\closeout\labelswrite  \input \jobname.lab\vfil\eject}
\let\labelend=\rel@x
\def\labelbegin#1\labelend{\setbox0=\vbox{\ialign{##\hfil\cr #1\crcr}}
     \MakeALabel }
\def\MakeFromBox{\gl@bal\setbox\FromLabelBox=\vbox{\Tenpoint
     \ialign{##\hfil\cr \the\sendername \the\FromAddress \crcr }}}
\def\MakeALabel{\vskip 1pt \hbox{\vrule \vbox{
        \hsize=\labelwidth \hrule\bigskip
        \leftline{\hskip 1\parindent \copy\FromLabelBox}\bigskip
        \centerline{\hfil \box0 } \bigskip \hrule
        }\vrule } \vskip 1pt plus 1fil }
\def\signed#1{\par \nobreak \bigskip \dt@pfalse \begingroup
  \everycr={\noalign{\nobreak
            \ifdt@p\vskip\signatureskip\gl@bal\dt@pfalse\fi }}%
  \tabskip=0.5\hsize plus \z@ minus 0.5\hsize
  \halign to\hsize {\strut ##\hfil\tabskip=\z@ plus 1fil minus \z@\crcr
          \noalign{\gl@bal\dt@ptrue}#1\crcr }%
  \endgroup \bigskip }
\newbox\letterb@x
\def\lettertext{\par \vskip\parskip \unvcopy\letterb@x \par }
\def\multiletter{\setbox\letterb@x=\vbox\bgroup
      \everypar{\vrule height 1\baselineskip depth 0pt width 0pt }
      \singlespace \topskip=\baselineskip }
\def\letterend{\par\egroup}
%
%
%
\newskip\frontpageskip
\newtoks\Pubnum   
\newtoks\Pubtype  \let\pubtype=\Pubtype
\newif\ifp@bblock  \p@bblocktrue
\def\PH@SR@V{\doubl@true \baselineskip=24.1pt plus 0.2pt minus 0.1pt
             \parskip= 3pt plus 2pt minus 1pt }
\def\PHYSREV{\papers\PhysRevtrue\PH@SR@V}

\def\titlepage{\FRONTPAGE\papers\ifPhysRev\PH@SR@V\fi
   \ifp@bblock\p@bblock \else\hrule height\z@ \rel@x \fi }
\def\nopubblock{\p@bblockfalse}
\def\endpage{\vfil\break}
\frontpageskip=12pt plus .5fil minus 2pt
\Pubtype={}
\Pubnum={}
\def\p@bblock{\begingroup \tabskip=\hsize minus \hsize
   \baselineskip=1.5\ht\strutbox \topspace-2\baselineskip
   \halign to\hsize{\strut ##\hfil\tabskip=0pt\crcr
       \the\Pubnum\crcr\the\date\crcr\the\pubtype\crcr}\endgroup}
\def\title#1{\vskip\frontpageskip \titlestyle{#1} \vskip\headskip }
\def\author#1{\vskip\frontpageskip\titlestyle{\twelvecp #1}\nobreak}

\def\address#1{\par\kern 5pt\titlestyle{\twelvepoint\it #1}}
\def\andaddress{\par\kern 5pt \centerline{\sl and} \address}

\def\abstract{\par\dimen@=\prevdepth \hrule height\z@ \prevdepth=\dimen@
   \vskip\frontpageskip\centerline{\fourteenrm ABSTRACT}\vskip\headskip }

%
%
%

\def\\{\rel@x \ifmmode \backslash \else {\tt\char`\\}\fi }
\def\sequentialequations{\rel@x \if\equanumber<0 \else
  \gl@bal\equanumber=-\equanumber \gl@bal\advance\equanumber by -1 \fi }
\def\journal#1&#2(#3){\begingroup \let\journal=\dummyj@urnal
    \unskip, \sl #1\unskip~\bf\ignorespaces #2\rm
    (\afterassignment\j@ur \count255=#3), \endgroup\ignorespaces }
\def\j@ur{\ifnum\count255<100 \advance\count255 by 1900 \fi
          \number\count255 }
\def\dummyj@urnal{%
    \toks@={Reference foul up: nested \journal macros}%
    \errhelp={Your forgot & or ( ) after the last \journal}%
    \errmessage{\the\toks@ }}

\def\topspace{\hrule height 0pt depth 0pt \vskip}

\def\Buildrel#1\under#2{\mathrel{\mathop{#2}\limits_{#1}}}
\def\becomes#1{\mathchoice{\becomes@\scriptstyle{#1}}
   {\becomes@\scriptstyle{#1}} {\becomes@\scriptscriptstyle{#1}}
   {\becomes@\scriptscriptstyle{#1}}}
\def\becomes@#1#2{\mathrel{\setbox0=\hbox{$\m@th #1{\,#2\,}$}%
        \mathop{\hbox to \wd0 {\rightarrowfill}}\limits_{#2}}}

\let\int=\intop         
\def\lsim{\mathrel{\mathpalette\@versim<}}
\def\gsim{\mathrel{\mathpalette\@versim>}}
\def\@versim#1#2{\vcenter{\offinterlineskip
        \ialign{$\m@th#1\hfil##\hfil$\crcr#2\crcr\sim\crcr } }}
\def\big#1{{\hbox{$\left#1\vbox to 0.85\b@gheight{}\right.\n@space$}}}
\def\Big#1{{\hbox{$\left#1\vbox to 1.15\b@gheight{}\right.\n@space$}}}
\def\bigg#1{{\hbox{$\left#1\vbox to 1.45\b@gheight{}\right.\n@space$}}}
\def\Bigg#1{{\hbox{$\left#1\vbox to 1.75\b@gheight{}\right.\n@space$}}}
\def\){\mskip 2mu\nobreak }
%
%
%
\let\sec@nt=\sec
\def\sec{\rel@x\ifmmode\let\n@xt=\sec@nt\else\let\n@xt\section\fi\n@xt}
\def\obsolete#1{\message{Macro \string #1 is obsolete.}}
\def\firstsec#1{\obsolete\firstsec \section{#1}}
\def\firstsubsec#1{\obsolete\firstsubsec \subsection{#1}}
\def\thispage#1{\obsolete\thispage \gl@bal\pagenumber=#1\frontpagefalse}
\def\thischapter#1{\obsolete\thischapter \gl@bal\chapternumber=#1}
\def\splitout{\obsolete\splitout\rel@x}
\def\prop{\obsolete\prop \propto }
\def\nextequation#1{\obsolete\nextequation \gl@bal\equanumber=#1
   \ifnum\the\equanumber>0 \gl@bal\advance\equanumber by 1 \fi}
\def\BOXITEM{\afterassigment\B@XITEM\setbox0=}
\def\B@XITEM{\par\hangindent\wd0 \noindent\box0 }
%
%
%
\def\phyzzx{PHY\setbox0=\hbox{Z}\copy0 \kern-0.5\wd0 \box0 X}
        
\everyjob{\xdef\today{\monthname~\number\day, \number\year}
        \input myphyx.tex }
\catcode`\@=12 
%

%
%
\catcode`\@=11 
\def\papersize{\hsize=40pc \vsize=53pc \hoffset=0pc \voffset=-2pc
   \advance\hoffset by\HOFFSET \advance\voffset by\VOFFSET
   \pagebottomfiller=0pc
   \skip\footins=\bigskipamount \normalspace }
\catcode`\@=12 
\papers
\vsize=23.cm
\hsize=15.cm

\Pubnum={NEIP-99-016 \cr
{\tt hep-th@xxx/9909220} \cr
September 1999}

\date={}
\pubtype={}
\titlepage
\title{{\bf Remarks on defining the DLCQ of quantum field theory\break 
 as a light-like limit}
}
\author{Adel~Bilal
\foot{Partially supported by the
Swiss National Science Foundation.}
}
\vskip .5cm
\address{
Institute of Physics, University of Neuch\^atel, CH-2000 Neuch\^atel, Switzerland \break
{\tt adel.bilal@iph.unine.ch}
}

\vskip 0.5cm
\abstract{The issue of defining discrete light-cone quantization  (DLCQ) in {\it field} 
theory as a light-like limit is investigated. This amounts to studying quantum field 
theory compactified on a space-like circle of vanishing radius in an appropriate 
kinematical setting. While this limit is unproblematic at the tree-level, it is 
non-trivial for loop amplitudes. In one-loop amplitudes, when the propagators are 
written using standard Feynman 
$\alpha$-parameters we show that, {\it generically}, in the limit of vanishing radius, 
one of the $\alpha$-integrals is replaced by a discrete sum and the (UV renormalized) 
one-loop amplitude has a finite light-like limit. This is analogous to what happens 
in string 
theory. There are however exceptions and the limit may diverge in certain theories 
or at higher loop order. We give a rather detailed analysis of the problems one might 
encounter. We show that quantum electrodynamics at one loop has a well-defined 
light-like limit.
}

%

\endpage
\pagenumber=1

\def\PL #1 #2 #3 {Phys.~Lett.~{\bf #1} (#2) #3}
\def\NP #1 #2 #3 {Nucl.~Phys.~{\bf #1} (#2) #3}
\def\PR #1 #2 #3 {Phys.~Rev.~{\bf #1} (#2) #3}
\def\PRL #1 #2 #3 {Phys.~Rev.~Lett.~{\bf #1} (#2) #3}
\def\CMP #1 #2 #3 {Comm.~Math.~Phys.~{\bf #1} (#2) #3}

\def\d{\delta}

\def\a{\alpha}
\def\b{\beta}
\def\o{\omega}
\def\m{\mu}
\def\n{\nu}

\def\l{\lambda}
\def\g{\gamma}
\def\f{\phi}
\def\rmd{{\rm d}}
\def\rd{\sqrt{2}}

\def\e{\epsilon}
\def\et{\tilde\epsilon}
\def\nabs{\vert n\vert}
\def\xm{x^-}
\def\xp{x^+}

\def\ap{\alpha'}
\def\to{\rightarrow}

\def\ro{R_0}
\def\ikd{\int{\rmd^d k\over (2\pi)^d}}
\def\ikdm{\int{\rmd^{d-1} k_\perp\over (2\pi)^{d-1}}}
\def\sp{{\sum_{n=0}^N}'}
\def\np{{\cal N}}
\def\psl{{p \hskip-1.8mm{/}}}

\REF\PH{S. Hellerman and J. Polchinski, {\it Compactification in the lightlike limit},  
{\tt hep-th/9711037}, \PR D59 1999 125002. }

\REF\SUSS{L. Susskind,  {\it Another conjecture about M(atrix) theory}, {\tt hep-th/9704080}.}

\REF\SEI{N. Seiberg, {\it Why is the matrix model correct?}, {\tt hep-th/9710009},
Phys. Rev. Lett. {\bf 7} (1997) 3577.}

\REF\SEN{A. Sen,  {\it D0-branes on $T^n$ and matrix theory}, {\tt hep-th/9709220},
Adv. Theor. Math. Phys. {\bf 2} (1998) 51.}

\REF\BILAL{A. Bilal, {\it A comment on compactification of M-theory on an 
(almost) light-like circle}, {\tt hep-th/9801047}, \NP B521 1998 202.} 

\REF\BILALII{A. Bilal, {\it DLCQ of M-theory as a light-like limit }, 
{\tt hep-th/9805070}, \PL B435 (1998) 312.}

{\bf \chapter{Introduction }}

Light-cone quantization is based on the idea
that one might use the light-cone coordinate $\xp\sim x^0+x^1$ as ``time" 
and the corresponding $P_+\sim P^-$ as hamiltonian. {\it Discrete} light cone quantization,
or DLCQ for short, in addition takes $\xm\sim x^0-x^1$ to be compact, i.e. to take values on
a circle of radius $\ro$: $\xm\simeq \xm +2\pi \ro$. One should note that the value of $\ro$
has no invariant meaning since the proper length of this circle is zero. The value of $\ro$
can be changed at will by a Lorentz transformation.

One may view this setup as resulting from a standard quantization with an ordinary time 
$t\equiv x^0$ and compactified space-like coordinate $x^1$ through an infinite Lorentz
boost. More precisely (see section 2) one starts with a space like circle of radius
$R=\e\ro$: $x^1\simeq x^1+2\pi\e\ro$ with small but finite $\e$. Through a large Lorentz
boost this is mapped to an almost light-like circle. In the $\e\to 0$ limit, the Lorentz
transformation becomes infinite and the circle truely light-like. It has been proposed [\PH]
to use this procedure as a definition for the DLCQ: Carry out the quantization of a given
theory compactified on a space-like circle of radius $\e\ro$. If the theory is Lorentz
invariant and if the $\e\to 0$ limit exists, the latter provides a clear definition of the
DLCQ of the same theory.

In particular this procedure should also provide a straightforward way to transpose the 
standard renormalisation into the DLCQ. This would be an advantage with respect to certain
current DLCQ treatments of QCD where renormalisation often looks a bit ad hoc.

Unfortunately, at present, it is not clear for which quantum theories this 
light-like  ($\e\to 0$) limit exists. Recent interest in this question arose in the context
of the DLCQ of M-theory [\SUSS,\SEI,\SEN]. However, it was pointed out in [\PH] that already
in the simple $\f^4$-theory the 4-point one-loop amplitude diverges, as $\e\to 0$, if no
momentum in the compact dimension is transferred across the loop. These authors [\PH]
advocated that this problem is generic, except in certain susy field theories.

Lateron it was shown that this problem does not occur in type II superstring 
theory and that there the $\e\to 0$ limit  exists, at least at one loop [\BILAL]. It was
argued that the same should also hold at higher loops and even non-pertubatively, and hence
probably also in M-theory [\BILALII]. It appeared from these studies that the mechanism did
not mainly rely on susy cancellations as might have been expected from [\PH] but was more
stringy in nature.

One might wonder whether the field theory limit ($\ap\to 0$) of the $\e\to 0$ 
limit of the string amplitude gives a finite result or whether some divergence appears.
Since the resulting field theories are highly supersymmetric and hence rather non-generic,
we have prefered to study general quantum field theories directly, trying to follow as much
as possible the string computation. It is well known  that a string one-loop amplitude
can be rewritten as a quantum field theory one-loop amplitude but with an infinite number of
particle species running around the loop. The string loop amplitude contains an integral
over the moduli $\nu_r$ of the punctures (localisations of the external propagators on the
torus). The field theory analogue of the complex $\nu_r$ are the real Feynman
$\a_r$-parameters. In the string computation [\BILAL] the $\e\to 0$ limit gives rise to a
complex $\delta$-function eliminating the integration over one of the moduli $\nu_r$,
replacing it by a finite sum. The resulting amplitude is finite except for precisely those
singularities required by unitarity.

We will show here that similarly in quantum field theory the $e\to 0$ limit gives rise to
a now real $\delta$-function. Generically, this $\delta$-function eliminates one of the
integrations over the Feynman parameters $\a_r$, replacing it by a finite sum, the resulting
expression having again only the singularities required by unitarity (after renormalisation).

This is the generic situation. If however {\it no} external momentum in the compact 
direction flows through the loop, then the argument of the $\delta$-function can be
vanishing, the $\delta(0)$ signalling a divergent $\e\to 0$ limit. This is precisely the
situation that was encountered in [\PH] for the $\f^4$-theory. In  $\f^3$-theory this may
not happen at one loop. Indeed, since the external lines must have non-vanishing momenta in
the compact direction,\foot{As will be explained in section 2, this is necessary in order to
correspond to finite energy states in the DLCQ.} for any theory with cubic vertices only
these momenta always flow through the loop. In particular, we show that the light-like limit
exists for QED at one loop. We also discuss some subtleties related to renormalisation.

At higher $L$-loop order the situation is less clear. Again, the generic situation 
is straightforward: there is one real $\delta$-function for each loop, eliminating $L$
integrals over  $\a_r$-parameters. This is quite encouraging for the string case of [\BILAL]
where the higher loop case is technically more involved. But we also  encounter 
non-generic situations (even in theories with cubic vertices only) 
where a loop subgraph inside a bigger loop has 
vanishing compact momenta on its external legs (which are just internal lines of the bigger
loop). In this case the $\e\to 0$ limit diverges. Although this is likely to ruin 
the existence of the light-like limit beyond one-loop, there might be cancellations 
that save it. Clearly a more detailed analysis
is needed beyond one loop to provide a clear answer, whether and for which theories the DLCQ
may be defined as a light-like limit to all orders in perturbation theory.

In section 2, we begin by setting up the relevant kinematic framework. In section 3 we 
study the scalar $\f^3$-theory in quite some detail, while in section 4 we extend these
results to quantum electrodynamics. Section 5 contains some discussion.

{\bf\chapter{Kinematics}}

We will start by considering a particle of mass $m$ descibed in a coordinate system
$(x^0,x^1,x^i)$ with  $i=2, \ldots d-1$, in a $d$-dimensional space-time with signature
$(+-\ldots -)$. $x^0$ is time and the spatial coordinate $x^1$ takes values on a circle of
radius $R=\e \ro$. All other spatial coordinates $x^i$, called transverse, as well as the
time $x^0$ are ordinary non-compact coordinates:
$$ x^0\simeq x^0 \quad , \quad  x^1 \simeq   x^1 + 2\pi \e R_0 \quad ,
\quad   x^i \simeq  x^i \ . 
\eqn\di$$ 
The momentum for an on-shell particle is
$$  p^1=-p_1={n\over \e\ro}\ ,\quad p^i \ {\rm arbitrary}\ , \quad
p^0=[p_1^2+p_i^2+m^2]^{1/2} \ge {\nabs\over\e\ro}
\ . 
\eqn\dii$$ 
At this stage $n$ is any integer, positive, negative or zero.

Now perform a Lorentz boost with boost parameter
$\b = {1-\e^2/2\over 1+\e^2/2}$. If $\e$ is small, this is a large boost. In the new
coordinate system    $\tilde x^\m$ it is convenient to define $\tilde x^\pm
=(\tilde x^0\pm \tilde x^1)/\rd$. Then simply $\tilde x^+={\e\over 2}(x^0+x^1),\ \tilde
x^-={1\over \e} (x^1-x^0)$ with periodicities
$$ \xm\simeq \xm +2\pi R_0 \quad , \quad
\xp\simeq\xp + \e^2 \pi R_0\ .
\eqn\diii$$
The corresponding transformed momenta are
$$\tilde p_+={1\over \e} (p_0+p_1)={p_0\over \e}-{n\over \e^2\ro} \quad , \quad
\tilde p_-={\e\over 2}(p_1-p_0)=-{\e\over 2}p_0-{n\over 2\ro} \ .
\eqn\div$$
Remember that $p_0\ge {\nabs\over \e\ro}$ and hence $\e p_0$ is ${\cal O}(1)$ while
${p_0\over \e}$ is ${\cal O}(1/\e^2)$, so that $\tilde p_-$ is ${\cal O}(1)$ and $\tilde
p_+$ is a priori ${\cal O}(1/\e^2)$.

Equation \diii\ shows that the light-cone coordinate $\tilde x^-$ is periodic as we want,
but also $\tilde x^+$  has a (small) periodicity which is unwanted. This can be eliminated
by a further coordinate redefinition (not a Lorentz transformation). Let
$$t=\tilde x^+-{\e^2\over 2}\tilde x^-\quad , \quad x^-=-\tilde x^- \quad \Rightarrow \quad
t\simeq t \quad , \quad x^-\simeq x^- -2\pi\ro \ .
\eqn\dv$$
Then $t$ is a non-periodic coordnate and $x^-$ has the desired periodicity. The metric is
$$\rmd s^2=2 \rmd t \rmd x^- -\e^2 (\rmd x^-)^2 -(\rmd x^i)^2
\eqn\dvi$$
so that the circle in the $x^-$ direction is not exactly light-like. It becomes truely
light-like in the limit $\e\to 0$. This is to be expected since the original circle \di\
has invariant length $2\pi\e\ro$ and a light-like circle must have zero invariant length.
The $\e\to 0$ limit gives the DLCQ setting. In the coordinates \dv\ the momenta are
$$ p_t=\tilde p_+={p_0\over \e} -{n\over \e^2\ro} \quad , 
\quad p_-=-\tilde p_- - {\e^2\over
2}\tilde p_+ = {n\over \ro} \ .
\eqn\dvii$$
Hence $p_-={n\over \ro}$ as expected in the DLCQ. 	

It is now easy to see that $n$ must be
positive if $m\ne 0$ and non-negative if $m=0$. Let first $n\ne 0$. Then, expanding the
square-root in \dii\ for very small $\e$: $p_0=p^0={\nabs\over \e\ro}+{\e\ro\over 2\nabs}(p_i^2+m^2)
+{\cal O}(\e^3)$ and hence
$$p_t={\nabs - n\over \e^2\ro} + {\ro\over 2\nabs}(p_i^2+m^2) +{\cal O}(\e^2) \ .
\eqn\dviii$$
Hence states with $n<$ have infinite DLCQ energy $p_t$ as $\e\to 0$, wile all state with
$n>0$ have finite $p_t$. Let now $n=0$. Then $p_0=(p_i^2+m^2)^{1/2}$ and $p_t={p_0\over
\e}$. Hence the only state with $n=0$ and finite DLCQ energy $p_t$ must have $m=p_i=0$,
i.e. is degenerate with the vacuum.

In the following, we will work in the coordinates \di, i.e. with a space-like circle of
radius $R=\e\ro$. In the $\e\to 0$ limit this is Lorentz equivalent to the DLCQ with radius
$\ro$. We have just seen that finite DLCQ energies for on-shell states require the
restriction $n>0$. Hence we will be interested in $\np$-point amplitudes in quantum field
theory with all external states having strictly positive momenta in the compact direction,
$p_{(r)}^1={n_r\over \e\ro}$ and study their $\e\to 0$ limit.

Possible divergencies can only occur in loop diagrams since the tree amplitudes are
entirely expressible in terms of scalar products of {\it on-shell} momenta $p_{(r)}\cdot
p_{(s)}$ and these are always finite as $\e\to 0$ (provided $n_r, n_s >0$):
$$p_{(r)}\cdot p_{(s)} = {n_s\over 2n_r} ((p_{(r)}^i)^2+m_r^2) +
{n_r\over 2n_s} ((p_{(s)}^i)^2+m_s^2) - p_{(r)}^i p_{(s)}^i + {\cal O}(\e^2) \ .
\eqn\dix$$

{\bf\chapter{Scalar quantum field theory : $\f^3$}}

The simplest quantum field theories to study are the scalar $\f^3$ or $\f^4$ theories. We
already know [\PH] that the simplest one-loop diagram in $\f^4$, namely the 4-point
amplitude, diverges as $\e\to 0$ if $n_1=n_2,\ n_3=n_4$. We will study instead $\l\f^3$
theory which does not present this pathology. In the course of our investigation we will
also better understand the origin of the problem of $\f^4$. The same problem will occur for
all $\f^k$ theories with $k\ge 4$. However, since many interesting theories, like QED, only
have cubic vertices it is useful to study the $\f^3$ theory in some detail.

{\bf\section{The 2-point one-loop amplitude}}

We will begin with a very detailed computation of the simplest one-loop diagram: the
two-point function. This will exhibit the basic mechanism of the $\e\to 0$ limit. If we 
call the external momentum $P$ and the loop momentum $k$ then the relevant self-energy
diagram is
$$ i\Pi(P)={1\over 2} \l^2 m^{4-d} \ikd {1\over (k^2-m^2) ((P-k)^2-m^2)}
\eqn\ti$$
where ${1\over 2}$ is the symmetry factor. We work with dimensional regularisation. (The
coupling constant $\l$ has dimension of mass as approriate in $d=4$.) Upon compactifying
$x^1$ on the circle of radius $\e\ro$ we have to replace $k^1 \to {n\over \e\ro}$ and
$\int\rmd k^1\to {1\over \e\ro}\sum_n$. We further do a Wick rotation to Euclidean
signature and denote $k_\perp\equiv (k^0, k^i)$. The external momentum is 
$P^1={N\over\e\ro}$ and $P_\perp$. Thus
$$ \Pi(P)={\l^2 m^{4-d}\over 4\pi\e\ro} \sum_n \ikdm 
{1\over \left[ \left( {n\over\e\ro}\right)^2 +k_\perp^2 +m^2\right]
\left[ \left( {N-n\over\e\ro}\right)^2 +\left( P_\perp-k_\perp\right)^2 +m^2\right]}
\ .
\eqn\tii$$
Note that although the external $N$ is positive, the  $n$ of the loop momentum has no a priori reason to be
restricted to positive values only. 

Now introduce an $\a$ parameter for each propagator, using ${1\over A_a}=\int_0^\infty
\rmd\a_a e^{-\a_a A_a}$ and then change variables $\a_1+\a_2 = \a$, ${\a_2\over \a}=\g$,
${\a_1\over \a}=1-\g$, so that after completing the squares in $n$ and $k_\perp$ we get
$$\eqalign{ 
\Pi(P)=&{\l^2 m^{4-d}\over 4\pi\e\ro} \int_0^\infty \rmd \a\ \a \int_0^1 \rmd \g
\sum_n \ikdm \ \times \cr
&\times \exp\left\{ -\a
\left[ \left( {n-\g N\over\e\ro}\right)^2 +\left(k_\perp-\g P_\perp\right)^2 
+\g(1-\g) \left( {N^2\over \e^2\ro^2}+P_\perp^2\right)  +m^2\right]
\right\}
\ . \cr
}
\eqn\tiii$$
Now, ${N^2\over \e^2\ro^2}+P_\perp^2$ is just $P^2$. 
We assume that $P^2$ is finite as $\e\to 0$.
Of course this is the case if $P$ is on shell, but we can consider more general cases.
\foot{
This point is slightly delicate: as long as we have Euclidean signature, $P^2$ is a 
sum  of positive terms, and if $(P^1)^2={N^2\over \e^2\ro^2}\to\infty$ there is no 
way $P^2$ can remain finite if all components are real. But we have to keep in mind 
that ultimately we are interested in Minkowski signature in which case $P^2$ is finite 
provided $P^0={N\over \e\ro} +{\cal O}(\e)$ as discussed in section 2.
}
The $k_\perp$ integration is trivially done after shifting the   integration
variables as usual:
$$ \Pi(P)={\l^2 m^{4-d}\over 2 (2\pi)^d} \int_0^\infty \rmd \a\ \a 
\left({\pi\over \a}\right)^{d-1\over 2}
\int_0^1 \rmd \g {1\over \e\ro}  \sum_n  
\exp\left\{ -\a
\left[ \left( {n-\g N\over\e\ro}\right)^2  +\g(1-\g) P^2  +m^2\right]
\right\}
\eqn\tiv$$
Exactly  as in [\BILAL] the $\e$-dependent part combines to give a $\d$-function:
$$ \lim_{\e\to 0} \ {1\over \e\ro} 
\exp\left\{ -\a \left( {n-\g N\over\e\ro}\right)^2 \right\}
= \left( {\pi\over \a}\right)^{1/2} \d (n-\g N)
\eqn\tv$$
so that
$$ \Pi(P)={\l^2 m^{4-d}\over 2 (2\pi)^d} \int_0^\infty \rmd \a\ \a 
\left({\pi\over \a}\right)^{d\over 2}
\int_0^1 \rmd \g   \sum_n  \d (n-\g N)
\exp\left\{ -\a  \left[ \g(1-\g) P^2  +m^2\right]  \right\}
\ .
\eqn\tvi$$

The $\e\to 0$ limit has produced a $\d$-function which eliminates one of the integrations
over the $\a$-parameters (here the one over $\g$) and replaces it by a discrete sum.
Indeed\foot{
Clearly, if $N=0$ one is in trouble. The problematic situation in $\f^4$ corresponds
effectively to the $N=0$ case with no external discrete momentum flowing through the loop.
Here however, $N>0$ and the problem does not arise.}
$\d (n-\g N)={1\over N} \d(\g-\g_n)$ with $\g_n={n\over N}$. However, only $n=0,\ldots N$
can contribute since only these $\g_n$ are within the interval of the $\g$-integration. The
momenta $k^1={n\over \e\ro}$ and $P^1-k^1={N-n\over \e\ro}$ of the loop propagators are
thus restricted to non-negative values. There is a slight subtlety here: $n=0$ and $n=N$
correspond to $\g=0$ and $\g=1$ which are just on the border of the integration interval.
With the convention $\int_0^a \d(x) f(x) \rmd x = {1\over 2} f(0)$ and similarly for the
upper bound, the values $n=0$ and $n=N$ should only contribute with an extra factor
${1\over 2}$ to the sum. We denote $\sp f(n) ={1\over 2} f(0) +{1\over 2} f(N)
+\sum_{n=1}^{N-1} f(n)$.

Finally, doing the trivial $\a$-integral and going back to Minkowski signature we get
$$ \Pi(P)={\l^2 m^{4-d}\over 2 (4\pi)^{d/2}} \Gamma\left({4-d\over 2}\right) 
{1\over N} \sp 
\left[ m^2 -  {n\over N} \left( 1-{n\over N}\right) P^2  \right]^{d-4\over 2}
\ .
\eqn\tvii$$
Note that the somewhat ambiguous terms $n=0$ and $n=N$ are independent of the external
momentum $P$ and thus can be absorbed into the mass renormalisation.

\subsection{Renormalisation in $d=4$}

In $d=4$, $\Pi(P)$ is logarithmically divergent. Setting $\et={4-d\over 2}$ we get
$$ \Pi(P)=-{\l^2 \over 2 (4\pi)^2} 
{1\over N} \sp \log
\left[ 1 -  {n\over N} \left( 1-{n\over N}\right) {P^2 \over m^2} \right]
-\d m^2
\eqn\tviii$$
with
$$ \d m^2= -{\l^2 \over 2 (4\pi)^2} \left( \Gamma(\et) +\log(4\pi) \right)  \ .
\eqn\tix$$
The infinite part of $\Pi(P)$ is $-\d m^2$ and is cancelled by a mass counterterm $+\d m^2$. 
We note that the
latter does {\it not} depend on $N$, i.e. is independent of the external momentum $P$ as
it should. Also note that $n=0$ and $n=N$ actually do not contribute to the finite part 
$\Pi(P)+\d m^2$.

We may shift any finite constant  between the finite part of $\Pi(P)$ and $\d m^2$. If we impose
the standard renormalisation condition $\Pi_{\rm R}(p^2=m^2)=0$ then it is easy to see that
$$ \Pi_{\rm R}(P)=-{\l^2 \over 2 (4\pi)^2} 
{1\over N} \sum_{n=1}^{N-1} \left\{ 
\log \left[ 1 -  {n\over N} \left( 1-{n\over N}\right) {P^2 \over m^2} \right]
- \log \left[ 1 -  {n\over N} \left( 1-{n\over N}\right)  \right]  \right\}
\ .
\eqn\tx$$
The terms $n=0$ and $n=N$ have disappeared from the sum. This means in particular that
after renormalisation the discrete components of the propagators in the loop are strictly
positive, just as is the case for external on-shell states. This is an important feature of
the DLCQ loop diagrams. However it appears below that  this seems not to hold for
higher-point one-loop diagrams in $\f^3$. Note that this implies $\Pi_{\rm R}(P)=0$ for
$N=1$.

From eq. \tx\ it is clear that the two-point function has branch cuts starting at
$P^2={N^2\over n(N-n)} m^2$ for $n=1, \ldots \left[ {N\over 2}\right]$. This is easily seen
to correspond to the threshold of production of two on-shell particles with
$p^1_{(1)}={n\over \e\ro}$ and $p^1_{(2)}={N-n\over \e\ro}$.

So far, all is satisfactory: the $\e\to 0$ limit exists, the mass counterterm does not
depend on $N$, the internal propagators have strictly positive $n$, and the renormalised
two-point function has the appropriate unitarity cuts.

\subsection{Renormalisation in $d=6$}

$d=6$ is the critical dimension of $\f^3$ beyond which it is not renormalisable, so it is
interesting to look at this case as well. Let now $\et={6-d\over 2}$. Then $\Pi(P)$
diverges quadratically and \tviii\ is replaced by
$$ \Pi(P)={\l^2 \over 2 (4\pi)^3} 
{1\over N} \sum_{n=1}^{N-1} 
\left[ 1 -  {n\over N} \left( 1-{n\over N}\right) {P^2 \over m^2} \right]
\log \left[ 1 -  {n\over N} \left( 1-{n\over N}\right) {P^2 \over m^2} \right]
-\d m^2 - (Z-1) P^2
\eqn\txi$$
with
$$\eqalign{
\d m^2 
&= {\l^2 \over 2 (4\pi)^3} \left( \Gamma(\et) +1 +\log(4\pi) \right)  \cr
(Z-1) 
&=  {(\l^2/m^2) \over 12 (4\pi)^3}  \left( \Gamma(\et) +1 +\log(4\pi) \right)
\left( 1-{1\over N^2}\right) \ . \cr
}
\eqn\txii$$
Again, $\d m^2$ does not depend on $N$, but $(Z-1)$ does.
Although this dependence disappears at
large $N$ as it should, it is likely that this $N$-dependence of the wave-function
renormalisation constant $Z$ signals some inconsistency. It is not clear to us whether and
how this could be resolved.

{\bf\section{$\np$-point one-loop amplitudes}}

The basic feature that has emerged from the above computation was the appearance of a
$\d$-fct whose argument involves a certain combination of $\a$-parameters and external
discrete momentum quantum numbers $n_r$, allowing to trivially perform one of the
$\a$-integrations. It is clear that the same will happen for any one-loop diagram with
cubic vertices only. The purpose of this subsection is to actually perform the 
calculation for an arbitrary number $\np$ of external legs in
the $\f^3$-theory and check that there are no hidden difficulties. We will give much less
details than in the previous subsection. We will be interested in 4 dimensions so that all
one-loop $\np$-point functions with $\np\ge 3$ are UV convergent. Otherwise, 
for $d=6$, one could
use dimensional regularisation for the 3-point function, all others being finite.

The main technical issue for the $\np$-point function is to find the most convenient change
of variables for the $\a$-parameters so that the $\d$-function can be used efficiently to
eliminate one integration and the resulting integrals are over a simple domain. The correct
change of variables is inspired from the string computation [\BILAL].

Let the external momenta be $p_r,\ r=1, \ldots \np$, all taken to be  incoming.
Momentum conservation then is $\sum_{r=1}^\np p_r=0$. Obviously, not all $p_r^1={n_r\over
\e\ro}$ then are positive, a negative $n_r$ just means that we are actually dealing with an
outgoing particle rather than an incoming. However, all $n_r$ are non-vanishing. The
momentum of the $r^{\rm th}$ propagator in the loop then is
$$ k_r=k-p_1-\ldots - p_{r-1}=k+ p_r+\ldots + p_\np \ .
\eqn\txiii$$
Using $\a$-parameters the product of the (Euclidean) propagators is
$$I_\np(k,p_r)=\int_0^\infty \ldots \int_0^\infty \prod_{r=1}^\np \rmd \a_r
\exp\left\{ -\sum_{r=1}^\np \a_r (k_r^2+m^2) \right\} 
\ .
\eqn\txiv$$
We change variables to
$$\b_i=\sum_{r=1}^i \a_r
\eqn\txv$$
and introduce the notation $\b_{ij}\equiv \b_i-\b_j$ which equals $\sum_{r=j+1}^i \a_r$ if
$i>j$, and $\b_{ij}=-\b_{ji}$. The Jacobian obviously is 1. One has
$$ \sum_{r=1}^\np \a_r k_r^2 = \left( \sum_{r=1}^\np \a_r \right) k^2
- 2 k \cdot \left( \sum_{r=1}^\np \a_r \sum_{i=1}^{r-1} p_i \right)
+  \sum_{r=1}^\np \a_r\left( \sum_{i=1}^{r-1} p_i \right)^2
\ .
\eqn\txvi$$
Now we have the following identities:
$$\eqalign{
\sum_{r=1}^\np \a_r \sum_{i=1}^{r-1} p_i = -\sum_{r=1}^\np \a_r \sum_{i=r}^\np p_i 
&=-\sum_{i\ge r} \a_r p_i = - \sum_{i=1}^\np \b_i p_i \ , \cr
\sum_{r=1}^\np \a_r \left(\sum_{i=1}^{r-1} p_i\right)^2 
= -\sum_{r=1}^\np \a_r \sum_{i=1}^{r-1} p_i \cdot \sum_{j=r}^\np p_j
&=-\sum_{i<r\le j} \a_r p_i\cdot p_j = - \sum_{i<j} \b_{ji}\, p_i\cdot p_j  \ , \cr
\left(\sum_i \b_i p_i\right)^2=-{1\over 2} \sum_{i,j} \b_{ij}^2\,  p_i\cdot p_j 
&= - \sum_{i<j} \b_{ji}^2\,  p_i\cdot p_j \cr
}
\eqn\txvia$$
where we have used momentum conservation. We can then rewrite eq. \txvi\ as
$$\sum_{r=1}^\np \a_r k_r^2 
= \b_\np \left( k+{1\over \b_\np} \sum_i \b_i p_i \right)^2
+ \sum_{i<j} \left( {\b_{ji}^2\over \b_\np} - \b_{ji} \right) p_i\cdot p_j 
\ .
\eqn\txvii$$
Rescaling the $\b_i$ as
$$ \b_\np = \a \quad , \quad {\b_i\over \b_\np} =\g_i \ , \ i=1, \ldots \np -1
\eqn\txviii$$
(with $\g_\np\equiv 1$ and $\g_{ij}\equiv\g_i-\g_j$) the product \txiv\ of the 
loop propagators becomes
$$\eqalign{
I_\np(k,p_r)=
& \int_0^\infty \rmd \a\ \a^{\np-1} \int_0^1 \rmd \g_{\np-1}
\int_0^{\g_{\np-1}} \rmd \g_{\np-2}\ldots \int_0^{\g_2} \rmd \g_1\ \times \cr
&\times \exp \left\{ -\a \left[ \left( k+\sum_i \g_i p_i \right)^2
+  \sum_{i<j} \left( \g_{ji}^2-\g_{ji}\right) p_i\cdot p_j 
+m^2 \right] \right\} \ . \cr
}
\eqn\txix$$

The domain of integraton of the $\g_i$ is 
$0\le \g_1\le \g_2 \le \ldots \le \g_{\np-1} \le 1$. If the integrand were symmetric under
permutation of any two $\g_i$ one could instead integrate over the full hypercube
$0\le \g_1,\g_2, \ldots  \g_{\np-1} \le 1$ (and divide by $\np !$). 
But permuting $\g_i$ and $\g_j$ does two
things: it permutes $p_i$ and $p_j$ and it changes the sign of the
$\g_{ji}$. The latter is easily fixed by writing $\vert\g_{ji}\vert$ instead. The former is
more interesting. We should really compute the $\np$-point one-loop Green's function. This
is obtained from a single diagram by adding all permutations of the external lines, i.e. by
adding all permutations of the $p_i$. To avoid overcounting, one has to keep one external
line (say $p_\np$) fixed, and moreover divide by 2 to avoid counting a diagram and its
mirror image twice. This Bose symmetrisation thus precisely amounts to extending the
integration range of the $\g_i, i=1,\ldots \np-1$ to the full hypercube. Thus the correct
Green's function is (not writing any coupling constants explicitly):
$$\eqalign{
G_\np(p_r)
&=\ikd {1\over 2} \sum_{{\rm permutations\ of}\atop p_1, \ldots p_{\np-1}} I_\np(k,p_r) \cr
&={1\over 2} \int_0^\infty \rmd \a\ \a^{\np-1} \int_0^1 \ldots \int_0^1
\prod_{r=1}^{\np-1} \rmd \g_r \ikd \times \cr
&\ \ \times \exp \left\{ -\a \left[ \left( k+\sum_i \g_i p_i \right)^2
+ {1\over 2} \sum_{i,j} \left( \g_{ji}^2-\vert\g_{ji}\vert\right) p_i\cdot p_j 
+m^2 \right] \right\} \ .  \cr
}
\eqn\txx$$

The sequel is straightforward. For a compact $x^1$ coordinate with radius $R=\e\ro$ one
again replaces $\int \rmd k^1 \to {1\over \e\ro} \sum_n$ and $k^1={n\over \e\ro}$, while
$p_i^1={n_i\over \e\ro}$. The integral over the $d-1$ other components of $k$ is trivial,
while the $\e$ dependent part gives\foot{
As before we assume that all $p_i\cdot p_j$ have  finite limits as $\e\to 0$. This is in
particular the case if all $p_i$ are on shell, cf eq. \dix.}
$$\lim_{\e\to 0}  {1\over \e\ro} 
\exp \left\{ -\a \left( {n+\sum_i \g_i n_i\over \e\ro} \right)^2 \right\}
=\left({\pi\over \a}\right)^{1/2} \d \left( n+\sum_i \g_i n_i \right) \ .
\eqn\txxi$$
Thus, after performing the $\a$-integration, we end up with
$$\eqalign{
G_\np(p_r)
={1\over 2 (4\pi)^{d/2}} \Gamma\left( \np -{d\over 2}\right)
&\int_0^1 \ldots \int_0^1 \prod_{r=1}^{\np-1} \rmd \g_r 
\sum_n \d \left( n+\sum_i \g_i n_i \right) \times \cr
&\times  \left[ m^2 
+ {1\over 2} \sum_{i,j} \left( \g_{ji}^2-\vert\g_{ji}\vert\right) p_i\cdot p_j
\right]^{{d\over 2}-\np}
\ .  \cr }
\eqn\txxii$$
It is straightforward to check that for $\np=2$ this coincides with eq. \tvi\ (after doing
the $\a$-integral, and up to the coupling constant we have dropped).

At this point we can use the $\d$-function to trivially do one of the $\g_r$-integrals. In
terms of the initial $\a_r$-parameters this would have been most complicated. We choose to
eliminate the $\g_{\np-1}$ integration. To do so, we introduce the notation $\{ x \}_f$,
meaning the fractional part of $x$, i.e. $\{ x \}_f=x+\tilde n$ for some 
$\tilde n\in {\bf Z}$ and 
$\{ x \}_f \in [0,1)$. Then $\g_{\np-1}$ only takes the following $\vert n_{\np-1}\vert$
discrete values
$$\g_{\np-1}^l  = {1\over \vert n_{\np-1}\vert}
\left(  \Big\{ - ( {\rm sign}\, n_{\np-1}) \sum_{i=1}^{\np-2} \g_i n_i \Big\}_f 
+ l \right) \quad , 
\quad l=0, \ldots  \vert n_{\np-1}\vert-1 
\ .
\eqn\txxiii$$
For $\np=2$ this simply gives $\g_i^l={l\over \vert n_1\vert}$ as before. Thus
$$\eqalign{
G_\np(p_r)
=&{1\over 2 (4\pi)^{d/2}} \Gamma\left( \np -{d\over 2}\right)
\int_0^1 \ldots \int_0^1 \prod_{r=1}^{\np-2} \rmd \g_r 
\cr
&\times  {1\over \vert n_{\np-1}\vert}
\sum_{l=0}^{\vert n_{\np-1}\vert -1}
\left[ m^2 
+ {1\over 2} \sum_{i,j} \left( \g_{ji}^2-\vert\g_{ji}\vert\right) p_i\cdot p_j
\right]^{{d\over 2}-\np}\Bigg\vert_{\g_\np=1,\ \g_{\np-1}=\g_{\np-1}^l}
\ .  \cr }
\eqn\txxiv$$
Although this final expression looks asymmetric between the $\np^{\rm th}$, the 
$(\np-1)^{\rm th}$ and the other external legs, it is clear from the derivation that it is
completely symmetric. This will also be obvious for the examples considered below.
It is a perfectly well-defined expression and we see that the
$\e\to 0$ limit exists for the $\np$-point amplitude under consideration. Of course,
$G_\np$ has those cuts required by unitarity. We have already seen this for $\np=2$.

As an explicit example one may take the triangle diagram, $\np=3$ in $d=4$. 
Recall that in this
case $\g_3= 1$. Let $\g_1\equiv \g$, so that 
$\g_2=\g_2^l=(\{ -\g\, n_1\, {\rm sign} n_2\}_f
+l)/\vert n_2\vert$ and
$$\eqalign{
G_{\np=3}(p_i)={1\over 32 \pi^2}\int_0^1 \rmd\g \sum_{l=0}^{\vert n_2\vert -1}
\Big[ &m^2 + p_1^2 \g(1-\g) +p_2^2\g_2(1-\g_2)\cr
& +p_1\cdot p_2 (\g+\g_2-\vert \g-\g_2\vert -2\g\g_2 ) \Big]^{-1} \ . \cr
}
\eqn\txxv$$
For fixed $l$, as $\g$ varies from 0 to 1, $\g_2^l$ will be discontinuous whenever $\g n_1$
is an integer, so that to actually evaluate the integral, the interval $[0,1]$ has to be
split into pieces. If we take e.g. $n_1=2,\ n_2=n_3=-1$ only $l=0$ is present and 
$\g_2=\{ 2\g \}_f$ so that we have
to split the interval into $[0,{1\over 2}]$ and $[{1\over 2},1]$. It is easy to see that
both pieces actually give the same contribution and, after Wick rotating back to Minkowski
signature, we get
$$G_{\np=3}^{n_1=2, n_2=n_3=-1}(p_i)={1\over 32\pi^2} \int_0^1 \rmd x
\left[ m^2-\left( {p_1^2\over 2} -p_2\cdot p_3\right) x + 
\left( {p_1^2\over 4} -p_2\cdot p_3\right) x^2 \right]^{-1}
\eqn\txxvi$$
where we used $p_2^2+p_1\cdot p_2=-p_2\cdot p_3$. This expression is manifestly symmetric
under exchange of $p_2$ and $p_3$ as it should since $n_2=n_3$. The integral is elementary
and can easily be done.\foot{
One gets ${1\over 32 \pi^2} (a+b)^{-1/2} 
\log \left\{\left[ 1+ \left( 1+{a\over b} \right)^{1/2}\right]\, / \, 
\left[ 1- \left( 1+{a\over b} \right)^{1/2}\right]\right\}$ with
$a=m^2(p_1^2-4m^2)$ and 
$b={p_1^2\over 2} -p_2\cdot p_3 -2m^2
= {1\over 4} [ (p_1^2-4m^2) + ((p_3-p_2)^2 - 4m^2) ]$.
} 
However, we already see from \txxvi\ that a branch cut will start
whenever the square bracket in the integrand is zero at the boundary of the integration
interval. At $x=0$ it is just $m^2\ne 0$, but vanishing at $x=1$ gives $p_1^2=4 m^2$,
giving the  branch cut as expected. Indeed, an incoming $n_1=2$ state can split into two
states with $n=1>0$ each. Note also that there are no cuts in $p_2^2$ or $p_3^2$ and this
is in agreement with the fact that $\vert n_2\vert = \vert n_3\vert=1$ cannot split into
two states both with $n>0$.

It is interesting to trace back what are the discrete components of the loop momenta 
that contribute. One of these is $n$ as entering the argument of 
$\delta( n+\sum_i \g_i n_i)$. When $\g_1\le \g_2\le \ldots$ this $n$ corresponds 
to the propagator between the first and the $\np^{\rm th}$ external line, but after 
Bose symmetrisation (any configuration of the $\g_i\le 1$) this can correspond to 
any of the internal propagators. The $\delta$-function implies 
$n=-\sum_i \g_i n_i = - n_\np -n_{\np-1} \g_{\np-1}^l -\sum_{i=1}^{\np-2} \g_i n_i$. 
For the above example with $\np=3$ and $n_1=2, n_2=n_3=-1$ this gives 
$n=1+\g_2-2\g=1 +\{ 2\g\}_f -2\g =1-[2\g]$, 
hence $n=1$ for $\g\in [0, {1\over 2})$ and $n=0$ for $\g\in [{1\over 2},1)$. We 
see that, contrary to what happened for the renormalised two-point function, 
internal propagators do occur with vanishing discrete momentum. In this example, 
it is clear from momentum conservation around the loop that one cannot have 
$n_1=2, n_2=n_3=-1$ with all discrete loop momenta strictly positive (for a 
given ``time slicing"). Another example is $\np=3$ with $n_1=4, n_2=n_3=-2$ 
and $\g_2^l=( \{4\g\}_f +l)/2$, $l=0,1$ and hence $n=2+l - [4\g]$ which takes 
the values $-1, 0, 1, 2, 3$.
The only way to evade the conclusion that internal propagators may have vanishing 
(or negative) discrete momentum would be to show that these diagrams actually vanish. 
But the above explicit example \txxvi\ tells us that this is not so.

{\bf\section{$\f^3$ beyond one loop}}

Beyond one loop the situation is less clear. Naively, we
proceed in exactly the same way, and expect that
the mechanism that produced the $\d$-function at one loop also operates here giving
one $\d$-function for each loop, so that $L$ $\a$-parameters will be
discretised. 
However, looking more carefully, we see that one important ingredient at one loop was 
that the external momenta are such that the $p_r\cdot p_s$ have finite limits as 
$\e\to 0$. Recall for example eq. \tiii\ for $\Pi(p)$ where we used that 
${N^2\over \e^2 \ro^2} +P_\perp^2=P^2$ is finite. If this self-energy diagram 
appears as an insertion in some {\it internal} line  with momentum 
$k$ of another loop, then we have $\Pi(k)$, i.e. $P\to k,\ N\to n$ and 
there is no reason why 
${n^2\over \e^2\ro^2}+k_\perp^2$ should be finite as $\e\to 0$. This does not mean 
that we cannot take the $\e\to 0$ limit but it might be different from what one 
naively expects. Let us proceed anyway.

Consider some $L$-loop diagram with $\np$ external lines. Then there are $I=\np+3L-3$
internal propagators. As before they are rewritten using $I$ $\a$-parameters. The exponent
then is some quadratic form in the $L$ loop momenta $k_i,\ i=1, \ldots L$, and in the
external momenta $p_r,\ r=1,\ldots \np$, with coefficients that are functions of the $\a_s$.
Just as one completes the square for $k$ at one loop, one successively completes the
squares for the $L$ $k_i$, starting with $k_1$, then for $k_2$, etc. For the non-compact
directions one then shifts the integration variables $k_i^\perp$ to $\tilde k_i^\perp$ and
does the integral trivially. For the compact direction $x^1$ one has $k_i^1={l_i\over
\e\ro}$ (analogue of $k^1={n\over \e\ro}$ at one loop) and
$\int \prod_{i=1}^L \rmd k_i^1 \to {1\over (\e\ro)^L} \sum_{l_1, \ldots l_L}$, 
as well as\foot
{
The definition of the $f_j^2$ will have to take into account terms like the above-mentioned $n^2$.
}
$$\lim_{\e\to 0} {1\over (\e\ro)^L} 
\exp \left\{ - \sum_{j=1}^L \left({f_j(l_i,\a_s)\over \e\ro}\right)^2 \right\} 
=\pi^{L\over 2} \prod_{j=1}^L \d\left( f_j(l_i, \a_s)\right) \ .
\eqn\txxvii$$
In practice one would like to find the most convenient change of variables $\a_s\to \g_s$
analogous to \txv, \txviii. As a matter of principle, however, this is not necessary. All
we need is a subset of $L$ $\a$'s (label them $\a_1,\ldots \a_L$) such that
$$\det \left( {\partial f_j(l_i,\a_s)\over \partial \a_i}\right)_{i,j=1, \ldots L}
\Bigg\vert_{f_j=0} \ne 0
\eqn\txxviii$$
so that we can solve the constraints imposed by the $\d$-functions in terms of the $\a_1,
\ldots \a_L$. 
Rather than trying to analyse this condition in general we consider a specific two-loop 
example. Take the one-loop two-point function $\Pi(P)$ computed above
with internal propagators having momenta $k$ and $P-k$. In one of the loop propagators, say
the one with momentum $k$, insert again this same $\Pi(k)$. The ``external" momenta of the
inserted sub-loop then simply are $k$ and $-k$. The discrete component of $k$ is
$k^1={n\over\e\ro}$. Suppose the problem mentioned above of diverging 
$k^2={n^2\over \e^2\ro^2}+k_\perp^2$ is somehow resolved. The condition \txxviii\ then 
in particular means that the $\d$-function appearing in $\Pi(k)$ can be used to 
eliminate the $\g$-integration appearing in $\Pi(k)$. We have seen that this is 
the case if $k^1={n\over \e\ro}\ne 0$ (now $n$ plays the role of $N$ above).
In the previous section, we have shown that in
the renormalised $\Pi_{\rm R}(P)$ only $n=1,\ldots N-1$ contribute (recall that
$P^1={N\over \e\ro}$). Thus $n\ne 0$, and the full (renormalised) two-loop diagram 
(probably) has a
well-defined finite limit as $\e\to 0$.
However, this example  does not  represent the generic situation: consider
inserting a one-loop two-point function $\Pi(k)$ into any of the internal propagators of the
triangle diagram. As we have discussed at the end of the previous section, 
in this case it appears the possibility that one of these internal
propagators has vanishing discrete momentum so that the inserted $\Pi(k)$ is taken at
$k^1=0$ and thus its $\e\to 0$ limit diverges. 

{\bf \chapter{QED}}

The purpose of this section is to show that the same mechanism that worked for the 
scalar $\f^3$ theory also works in a realistic theory containing fermions and having 
gauge invariance.

We will explicitly study the behaviour of two UV divergent one-loop 
diagrams in QED in four
dimensions, namely the vacuum polarisation and the electron self-energy. 
Then we will comment on the other one-loop and on higher-loop diagrams. As
before, we work with dimensional regularisation performed on the Wick rotated Euclidean
integrals.

\sectionnumber=0
{\bf\section{ Vacuum polarisation}}

The vacuum polarisation diagram $\o_{\m\n}(p)$ has a superficial degree of divergence 2,
but it is well-known that $\o_{\m\n}(p)=i(p_\m p_\n- g_{\m\n} p^2) \o(p)$ with $\o(p)$ only
logarithmically divergent. In the present case, the discrete component is $p_1$ and we must
treat $\o_{11}, \o_{1i}$ and $\o_{ij}\ (i,j=0,2,3)$ separately. We begin with the Wick
rotated expression 
$$\o_{\m\n}(p)=i e^2 m^{4-d} \ikdm  {1\over 2\pi\e\ro}\sum_n
{
k_\m(k-p)_\n + k_\n(k-p)_\m - \d_{\m\n}\left( k\cdot(k-p)+m^2\right) 
\over
(k^2+m^2) \left( (k-p)^2+m^2\right)
}
\eqn\qi$$
with $m$ being the mass of the charged fermion (electron). 
We denote $p^1={N\over \e\ro}$ as
before and $k^1={n\over \e\ro}$. Introducing $\a$-parameters and going through the same
kind of algebra as for the self-energy diagram in the $\f^3$ theory, 
we get for $\o_{1i}$ in the $\e\to 0$ limit
$$\eqalign{
\o_{1i}(p)=-{i e^2 m^{4-d}\over (4\pi)^{d/2}}
&\int_0^\infty \rmd \a\ \a^{1-{d\over 2}} \int_0^1 \rmd \g \sum_n
\left( {n\over \e\ro}(1-\g) p_i + {N-n\over \e\ro} \g p_i \right) \times
\cr
&\times \d(n-\g N) 
\exp\left\{ -\a \left[ m^2+\g(1-\g) p^2\right]\right\} \ . \cr
}
\eqn\qii$$
Using $\d(n-\g N)$ to replace $n$ by $\g N$ in the parenthesis we can extract a factor
${N\over \e\ro} p_i = p_1 p_i$ so that indeed $\o_{1i}=i p_1 p_i\, \o(p)$ with
$$\eqalign{
\o(p)
&=-{ 2 e^2 m^{4-d}\over (4\pi)^{d/2}}
\int_0^\infty \rmd \a\ \a^{1-{d\over 2}} \int_0^1 \rmd \g \sum_n
\g(1-\g)  \d(n-\g N) \exp\left\{ -\a \left[ m^2+\g(1-\g) p^2\right]\right\}  \cr
&={e^2\over 8\pi^2} {1\over N} \sum_{n=1}^{N-1} {n\over N} \left( 1-{n\over N} \right)
\log\left[ 1+{n\over N}\left( 1-{n\over N} \right) {p^2\over m^2} \right] \ 
+ \ (Z_3-1) \cr
&\equiv \o_{\rm R}(p) + \ (Z_3-1) \cr
}
\eqn\qiii$$
with
$$\eqalign{
(Z_3-1)&=-{e^2\over 8\pi^2} 
{1\over N} \sum_{n=1}^{N-1} {n\over N} \left( 1-{n\over N} \right)
\left[ \Gamma(\et) + \log 4\pi \right] \cr
&=-{e^2\over 48\pi^2} \left( 1- {1\over N^2}\right)
\left[ \Gamma(\et) + \log 4\pi \right] \ . \cr
}
\eqn\qiv$$
We note two things. Again, the terms $n=0$ and $n=N$ do not contribute to $\o(p)$ or
$Z_3-1$. Again also, the counterterm $Z_3-1$  explicitly depends on $N$. As in the $\f^3$
theory, the first fact is encouraging, while the second is puzzling. Let's compare with the
standard (Euclidean) result in the non-compact case. There
$$\o_{\rm non-compact}(p) ={e^2\over 8\pi^2} \int_0^1 \rmd\g\ \g (1-\g)
\log\left[ 1+\g (1-\g) {p^2\over m^2}\right] + (Z_3-1) \ .
\eqn\qv$$
So all that has happened is $\g\to {n\over N}$ and 
$\int_0^1 \rmd\g \to {1\over N} \sum_{n=1}^{N-1}$.

We still need to check that $\o_{11}(p)=-i p_\perp^2 \o(p)$ and 
$\o_{ij}(p)=i (p_i p_j -\d_{ij} p^2)\o(p)$. Here we encounter an interesting subtlety. In
both cases the numerator of \qi\ contains a term $k_1(k-p)_1={n(n-N)\over \e^2\ro^2}$ so
that the $\e$-dependent terms are of the form
$${1\over \e\ro} \left[ a\ {n(n-N)\over \e^2\ro^2} + b \right] 
\exp\left\{ -\a \left( {n-\g N\over \e\ro}\right)^2 \right\} \ .
\eqn\qvi$$
In this case one has to develop in $\e$ to next to leading order:
$${1\over \e\ro} \exp\left\{ -\a \left( {n-\g N\over \e\ro}\right)^2 \right\}
\ \sim\ 
\left( {\pi\over \a}\right)^{1\over 2} \left[ \d(n-\g N) +{\e^2\ro^2\over 4 \a}
\d''(n-\g N) + {\cal O}(\e^4) \right]
\eqn\qvii$$
as can easily be seen by multiplying by a test function $f(n)$  and integrating $\int \rmd
n$ (or equivalently by using a test function $f(\g)$ and integrating $\int \rmd \g$). As a
result, as $\e\to 0$, expression \qvi\ gives
$$\left( {\pi\over \a}\right)^{1\over 2} 
\left[ a\ {n(n-N)\over \e^2\ro^2} + b + {a\over 2\a} \right]  \d(n-\g N) \ .
\eqn\qviii$$
The extra contribution $\sim {a\over 2\a}$ is crucial. The rest of the computation is
straightforward algebra, and we indeed find that 
$\o_{\m\n}(p)=i(p_\m p_\n -\d_{\m\n} p^2) \o(p)$ for all $\m,\n$.

{\bf\section{Fermion self-energy}}

The computation is straightforward and the result is again identical to the standard result
of the non-compact case but with the replacement $\g\to {n\over N}$, $\int_0^1\rmd\g \to
\sp$. This time the terms $n=0$ and $n=N$ contribute to the regularised self-energy and
both terms must be included with a factor ${1\over 2}$ into the sum. Then
$$\eqalign{
\Sigma(p)={e^2 m^{4-d}\over (4\pi)^{d/2}} \Gamma\left( {4-d\over 2}\right)
&{1\over N} \sp \left[ d m - (d-2) \left(1-{n\over N}\right) \psl \right] \times\cr
&\times \left[ {n\over N}\left(1-{n\over N}\right) p^2 + {n\over N} m^2 
+ \left(1-{n\over N}\right) \m^2 \right]^{d-4\over 2}\cr
}
\eqn\qix$$
where $\m^2$ is an IR regulator (photon mass) needed to keep the $n=0$ term finite. We have
$$\eqalign{
\Sigma(p)
=&-{e^2 \over (4\pi)^2} 
{1\over N} \sp \left[ 4 m - 2 \left(1-{n\over N}\right) \psl \right]
\log\left[ {n\over N}\left(1-{n\over N}\right) {p^2\over m^2} + {n\over N}
+ \left(1-{n\over N}\right) {\m^2\over m^2} \right]\cr
&+\d m + (Z_2-1) (\psl -m)  \cr
}
\eqn\qx$$
with 
$$\eqalign{
\d m&={e^2\over (4\pi)^2} \left( 3\Gamma(\et) + 3 \log 4\pi -1 \right) m \cr
(Z_2-1) &=-{e^2\over (4\pi)^2}\left( \Gamma(\et)+\log 4\pi -1\right) 
\ . \cr
}
\eqn\qxi$$
We note that  $\d m$ and
$(Z_2-1)$ are independent of $N$.
We also note that the terms $n=0$ or $n=N$ in $\Sigma(p)$ are $\sim m$ or $\sim \psl$. They
can be eliminated from the  renormalised self-energy since we may still change 
$\Sigma_{\rm R}$ by finite counterterms. In particular, if we impose the standard
renormalisation condition $\Sigma_{\rm R}(\psl=m)=0$ and ${\partial \Sigma_{\rm R}\over
\partial \psl} (\psl=m)=0$ we find
$$\eqalign{
\Sigma_{\rm R}(p)
=-{e^2 \over (4\pi)^2} 
{1\over N} \sum_{n=1}^{N-1} 
&\Bigg\{ \left[ 4 m - 2 \left(1-{n\over N}\right) \psl \right]
\log\left[ {N\over n} + \left( {N\over n}-1\right) {p^2\over m^2}\right] \cr
&+ 4\left( {N\over n}-{n\over N}\right) (\psl -m) \Bigg\}  \cr
}
\eqn\qxii$$
with no contributions from $n=0$ or $n=N$ ! Note that we have set $\m=0$ because it was
only needed for the $n=0$ term which is no longer present.

{\bf\section{Other one-loop and higher loop diagrams}}

While the other one-loop diagrams (vertex function, etc) are technically more involved, 
in particular due to the $\g$-matrix algebra, there are no new difficulties to be 
encountered\foot
{
A minor complication arises when one wants to extend the integration domain of the 
$\g_i$ from $0\le\g_1\le \g_2 \le \ldots \le \g_{\np -1}\le 1$ to the full 
hypercube $0\le \g_i\le 1$. In the scalar $\f^3$ theory this was achieved by 
Bose symmetrisation. Here, in QED one has to separately Bose symmetrise external 
photon lines, and separately antisymmetrise incoming and outgoing fermion lines. 
As a result, we do not get the same type of compact expression with the $\g_i$ all 
integrated over the hypercube. This is easy to see for the vertex function. 
Nevertheless, one can still use the $\d$-function to eliminate one of the $\g_i$ 
which is replaced by a finite sum, although the lower and upper bound of the sum 
now depend on the other integration variable(s). In the case of the vertex 
function one can very explicitly evaluate the diagram in the $\e\to 0$ limit, 
which is indeed  finite, and study its properties.
}
and we can safely conclude that all one-loop amplitudes in QED have a finite 
light-like limit. As in $\f^3$-theory, the situation beyond one loop is less clear.

{\bf \chapter{Conclusions}}

We have studied the issue of defining discrete light-cone quantization  in {\it field} 
theory as a light-like limit. While this limit is unproblematic at the tree-level, it is 
non-trivial for loop amplitudes. We have seen that in theories with cubic vertices only, 
and in particular in QED, all one-loop amplitudes have a finite and well-defined 
light-like limit (provided the external states have non-vanishing discrete momenta which 
is the appropriate kinematic setting  corresponding to the DLCQ). Moreover, we have seen 
that these amplitudes have exactly the cuts required by unitarity. When they are 
written using standard Feynman $\alpha$-parameters, we showed that the only thing 
that happens in the light-like limit is the replacement of
one of the $\alpha$-integrals  by a discrete sum.  This is analogous to what happens 
in string 
theory. These  one-loop amplitudes are actually often easier to compute explicitly 
than the standard ones since they involve one less integral.

There are however some puzzling features: we have seen in the examples of the 
scalar self-energy (in $d=6$) and the QED vacuum polarisation 
diagrams that certain counterterms depend explicitly on $N$, i.e. 
on the discrete component of momentum of the external particle. It is not clear 
to us whether and how this might make sense. Also, while for the self-energy 
diagrams all internal discrete momenta are strictly positive, this is not always 
the case for the higher-point functions (vertex function, etc).

Related to this last point, we have argued that beyond one loop we are probably 
going to encounter problems and the light-like limit is likely to diverge, 
although a more detailed analysis is needed to settle the question. It should 
be noted that the troublesome situation occurring at two and more loops is  
very specifically field theoretic and does not affect the conjectured existence of a 
light-like limit of string theory to all orders in perturbation theory [\BILALII].

\ack
This work was partially supported by the Swiss National Science Foundation.

\refout

\end